\documentclass{article}
\usepackage[T1]{fontenc}
\usepackage{graphics}

\makeatletter

\providecommand{\LyX}{L\kern-.1667em\lower.25em\hbox{Y}\kern-.125emX\@}

\makeatother

\begin{document}

\title{Lattice QCD Calculations of\\
the Sigma Commutator}

\author{D. B. Leinweber\thanks{dleinweb@physics.adelaide.edu.au}~,
A. W. Thomas\thanks{athomas@physics.adelaide.edu.au}~, and 
S. V. Wright\thanks{swright@physics.adelaide.edu.au}\\
{\small\it Department of Physics and Mathematical Physics}\\
{\small\it and Special Research Centre for the Subatomic Structure of
Matter,}\\
{\small\it University of Adelaide, Adelaide 5005, Australia}}

\maketitle

\vspace{-7.cm}
\hfill ADP-00-01/T390
\vspace{7.cm}

\begin{abstract}
As a direct source of information on chiral symmetry breaking within
QCD, the sigma commutator is of considerable importance. With recent
advances in the calculation of hadron masses within full QCD it is of
interest to see whether the sigma commutator can be calculated directly
from the dependence of the nucleon mass on the input quark mass. We show
that provided the correct chiral behaviour of QCD is respected in the
extrapolation to realistic quark masses one can indeed obtain a fairly
reliable determination of the sigma commutator using present lattice data.
Within two-flavour, dynamical-fermion QCD the value obtained lies in the
range 45 to 55 MeV.
\end{abstract}

In the quest to understand hadron structure within QCD, small violations 
of fundamental symmetries play a vital role. The sigma commutator,
\( \sigma _{N} \):
\begin{equation}
\label{eqn:Full-Commutator}
\sigma _{N}=\frac{1}{3}\left\langle N\right| \left[ Q_{i5},\left[ Q_{i5},\cal {H}\right] \right] \left| N\right\rangle 
\end{equation}
(with \( Q_{i5} \) the two-flavour (\( i \)=1, 2, 3) axial charge) is an 
extremely
important example. Because \( Q_{i5} \) commutes with the QCD Hamiltonian in
the chiral SU(2) limit, the effect of the double commutator is to pick 
out the light
quark mass term from \( \cal {H} \):
\begin{equation}
\label{eqn:SU(2)-Commutator}
\sigma _{N}=\left\langle N\right| \left( m_{u}\bar{u}u+m_{d}\bar{d}d\right) \left| N\right\rangle 
\end{equation}
Neglecting the very small effect of the \( u-d \) mass difference we can write
Eq.~(\ref{eqn:SU(2)-Commutator}) in the form
\begin{eqnarray}
\sigma _{N} & = & \left\langle N\right| \bar{m}\left( \bar{u}u+\bar{d}d\right) \left| N\right\rangle \\
\label{eqn:SME}
 & = & \bar{m}\frac{\partial M_{N}}{\partial \bar{m}} 
 \label{eqn:FH-Commutator} 
\end{eqnarray}
with \( \bar{m}=(m_{u}+m_{d})/2 \). Equation (\ref{eqn:FH-Commutator}) follows
from the Feynman-Hellman theorem \cite{FeynmanHellman}.

While there is no direct experimental measurement of \( \sigma _{N} \), the
value inferred from world data has been \( 45\pm 8 \) MeV \cite{Sainio} for
some time. Recently there has been considerable interest in this value
because of progress in the determination of the pion-nucleon scattering
lengths \cite{PSI,LET} and new phase shift analyses 
\cite{Arndt,Bugg}. For an excellent summary of the sources of the
proposed variations and the disagreements between various investigators
we refer to the excellent review of Kneckt \cite{Kneckt}. For our
purposes the experimental value is of limited interest as the full
lattice QCD calculations upon which our work is based involve only two
active flavours. Nevertheless, as a guide, the current work suggests that
the best value of $\sigma_{N}$ may be between 8 and 26 MeV larger than
the value quoted above \cite{Kneckt}.

Numerous calculations of \( \sigma _{N} \) have been made within QCD
motivated models \cite{models} and there has been considerable work
within the framework of chiral perturbation theory \cite{ChPT}.
However, direct calculations of $\sigma_{N}$ within QCD itself have
proven to be difficult.  Early attempts \cite{CABASINO} to extract
$\sigma_N$ from the quark mass dependence of the nucleon mass in
quenched QCD (using Eq.(\ref{eqn:FH-Commutator}) ) 
produced values in the range 15 to 25 MeV.  Attention
subsequently turned to determining $\sigma_N$ by calculating the
scalar matrix element of the nucleon 
$\langle N|\bar u u + \bar d d|N\rangle$.
There it was discovered that the sea quark loops
make a dominant contribution to $\sigma_N$ \cite{LIU,FUKUGITA}.  
These works, based on quenched QCD simulation, found values in
the 40 to 60 MeV range, which are more compatible with the
experimental values quoted earlier.

On the other hand, the  most recent estimate of $\sigma_N$, 
and the only one based on a two-flavour,
dynamical-fermion lattice QCD calculation, comes from the SESAM collaboration.
They obtain a value of $18 \pm 5$ MeV \cite{SESAM}, through a direct
calculation of the scalar matrix element 
$\langle N|\bar u u + \bar d d|N\rangle $.
The discrepancy from the
quenched results of Refs.\ \cite{LIU,FUKUGITA} is not so much an
unquenching effect in the scalar matrix element but rather a
significant suppression of the quark mass in going from quenched to
full QCD.  The difficulty in all approaches which evaluate 
$\langle N|\bar u u + \bar d d|N\rangle $ is that neither it nor $\bar m$
is renormalization group invariant. One must reconstruct 
the scale invariant result from the product of 
the scale dependent matrix element and the scale dependent 
quark masses.  The latter are extremely difficult to determine
precisely and are the chief source of uncertainty in this approach.

An additional difficulty in extracting $\sigma_{N}$ from lattice
studies is the need to extrapolate from quite large pion masses,
typically above 500 or 600 MeV.  An important innovation adopted by
Dong {\it et al.}, but not by the SESAM collaboration, was to
extrapolate the computed values of 
$\langle N| \bar u u + \bar d d |N\rangle $ using
a form motivated by chiral symmetry, namely $a + b \bar m
^{\frac{1}{2}}$.  On the other hand, the value of $b$ used was {\em
not} constrained by chiral symmetry and higher order terms of the
chiral expansion were not considered.  Furthermore, since the work was
based on a quenched calculation, the chiral behaviour implicit in the
lattice results involves incorrect chiral coefficients \cite{Sharpe}.

Our work is motivated by recent, dramatic improvements in computing
power which, together with the development of improved actions
\cite{ImprovedActions}, mean that we now have accurate calculations of
the mass of the nucleon within {\em full QCD} (for two flavours) as a
function of \( \bar{m} \) down to \( m_{\pi }\sim 500 \) MeV. (Since
\( m_{\pi }^{2} \) is proportional to \( \bar{m} \) over the range
studied we choose to display all results as a function of \( m_{\pi
}^{2} \).) In addition, CP-PACS has recently published a result at \(
m_{\pi }\sim 300 \) MeV , albeit with somewhat large errors. Provided
that one has control over the extrapolation of this lattice data to
the physical pion mass, \( m_{\pi }\equiv \mu =140 \) MeV, one can
calculate \( \sigma _{N} \) by evaluating
Eq.~(\ref{eqn:FH-Commutator}) at the physical pion mass. Note that this
approach has the important advantage over the calculation of the scalar
density that one {\em only} needs to work with renormalization group
invariant quantities. We therefore
turn to a consideration of the method of extrapolation.

\begin{figure}[hbt]
{\par\centering \resizebox*{0.95\textwidth}{!}{\rotatebox{90}{\includegraphics{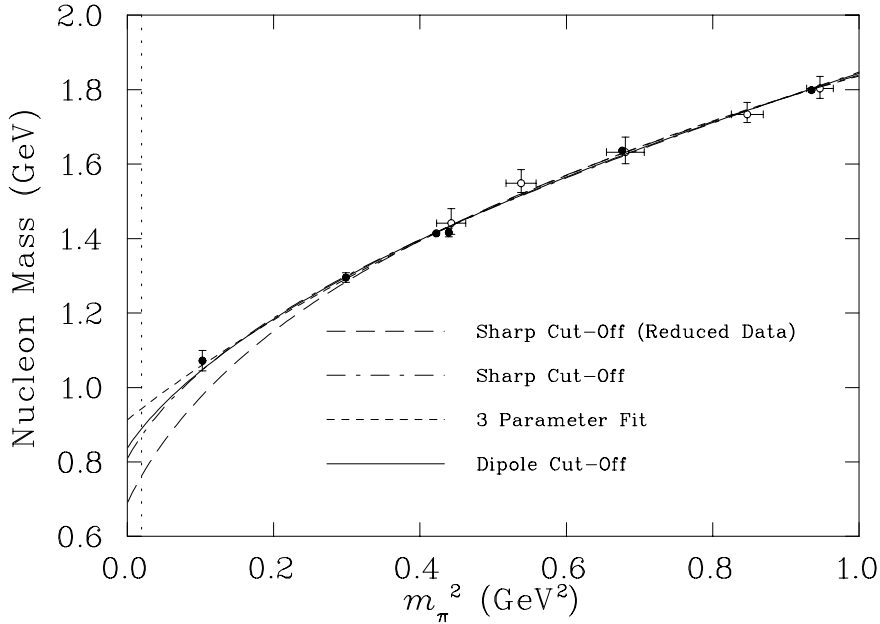}}} \par}
\caption{Nucleon mass calculated by CP-PACS (solid points) 
and UKQCD (open points),
as a function of \protect\( m_{\pi }^{2}\protect \), both are scaled by 5\% to
improve consistency. The solid curve is a fit to
Eq.~(\ref{eqn:Full-Eqn}) with a 1.225 GeV dipole form factor, while
the dashed curve is the same fit using a sharp cut-off form factor
(\protect\( \theta \protect \)-function). The short-dash curve is a
fit to Eq.~(\ref{eqn:Phen_Form_CP-PACS}), and the long-dash curve is a
fit to Eq.~(\ref{eqn:Full-Eqn}) {\em excluding} the lowest data
point. The vertical line indicates the physical pion mass.
\label{fig:Fits to Data}} 
\end{figure}

The lattice data for the nucleon mass calculated by UKQCD
\cite{Allton:1998gi} and CP-PACS \cite{Aoki:1999ff} is shown in
Fig.~\ref{fig:Fits to Data}. Both groups cite a 10\% uncertainty in
setting the lattice scale, so we have scaled the former down and the
latter up by 5\% so that the data sets are consistent.  Over almost
the entire range of \( m_{\pi }^{2} \), the data shows a dependence on
quark mass that is essentially linear. However, the preliminary point
at \( m_{\pi }^{2}\sim 0.1 \) GeV\( ^{2} \) suggests some curvature in
the low mass region. This is indeed expected on the basis of chiral
symmetry with the leading non-analytic (LNA) correction (in terms of
\( \bar{m} \)) being proportional to \( m_{\pi }^{3} \) (\(
\bar{m}^{3/2} \)):
\begin{equation}
\label{eqn:LNA-term}
\delta M_{N}^{\mbox {\tiny LNA}}=\gamma ^{\mbox {\tiny LNA}}m_{\pi }^{3}\, ,\, \, \gamma ^{\mbox {\tiny LNA}}=\frac{-3g_{A}^{2}}{32\pi f_{\pi }^{2}}\, .
\end{equation}
These observations led the CP-PACS group to extrapolate their data with the
simple, phenomenological form:
\begin{equation}
\label{eqn:Phen_Form_CP-PACS}
M_{N}=\tilde{\alpha }+\tilde{\beta }m_{\pi }^{2}+\tilde{\gamma }m_{\pi }^{3}\, .
\end{equation}
The corresponding fit to the combined data set, using
Eq.~(\ref{eqn:Phen_Form_CP-PACS}), is shown as the short-dashed curve
in Fig.~\ref{fig:Fits to Data} and the parameters
$(\tilde{\alpha},\tilde{\beta},\tilde{\gamma}) = (0.912,1.69,-0.761)$
(the units are appropriate powers of GeV). This yields a value for the
sigma commutator, \( \sigma _{N}^{(p)}=29.7 \) MeV, where the
superscript stands for ``phenomenological''.

The difficulty with this purely phenomenological 
analysis was discussed in Ref.~\cite{Leinweber:1999ig}.
That is, the value of \( \tilde{\gamma }=-0.761 \) is 
almost an order of magnitude
smaller than the model independent LNA 
term, \( \gamma ^{\mbox {\tiny LNA}}=-5.60 \)
GeV\( ^{-2} \). Clearly this presents some concern when evaluating \( \sigma _{N} \),
because of the derivative required.
\begin{figure}[hbt]
{\par\centering \includegraphics{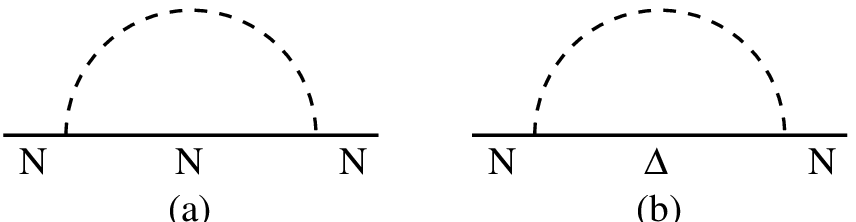} \par}
\caption{One-loop pion induced self energy of the nucleon.\label{fig:SE-N?N}}
\end{figure}
An alternative approach to this problem was recently suggested by
Leinweber et al. \cite{Leinweber:1999ig}.  They realised that the 
pion loop diagrams, Fig.~\ref{fig:SE-N?N}(a)
and \ref{fig:SE-N?N}(b) not only yield the most 
important non-analytic structure
in the expression for the nucleon mass, but amongst all the possible
meson baryon states which contribute to the nucleon mass within QCD,
they alone give rise to a significant variation of the nucleon mass as
$m_{\pi} \rightarrow 0$. 
In Ref.~\cite{Leinweber:1999ig} it was suggested that one should
extrapolate \( M_{N} \) as a function of quark mass using:
\begin{equation}
\label{eqn:Full-Eqn}
M_{N}=\alpha +\beta m_{\pi }^{2}+\sigma _{NN}(m_{\pi },\Lambda )+\sigma _{N\Delta }(m_{\pi },\Lambda )\, ,
\end{equation}
where \( \sigma _{NN} \) and \( \sigma _{N\Delta } \) are the
self-energy contributions of Figs.~\ref{fig:SE-N?N}(a) and
\ref{fig:SE-N?N}(b), respectively, using a sharp cut-off in momentum,
\( \theta (\Lambda -k) \).  The full analytical expressions for \(
\sigma _{NN} \) and \( \sigma _{N\Delta } \) are given in
Ref.~\cite{Leinweber:1999ig}. For our purposes it suffices that they
have precisely the correct LNA and next-to-leading non-analytic
behaviour required by chiral perturbation theory as \( m_{\pi
}\rightarrow 0 \). In addition, \( \sigma _{N\Delta } \) contains the
correct, square root branch point 
(\( \sim [m_{\pi }^{2}-(M_{\Delta}-M_{N})^{2}]^{\frac{3}{2}} \))
at the \( \Delta -N \) threshold, which is essential for
extrapolations from above the \( \Delta -N \) threshold.

{}Fitting Eq.~(\ref{eqn:Full-Eqn}) to the data, including the point near 0.1
GeV\( ^{2} \), gives the dot-dash curve 
in Fig.~\ref{fig:Fits to Data} (\( (\alpha ,\beta ,\Lambda )
=(1.42,0.564,0.661) \)).
The corresponding value of \( \sigma _{N} \) is 54.6 MeV 
and the physical nucleon
mass is 870 MeV. Omitting the lowest data 
point from the fit yields the long-dash
curve in Fig.~\ref{fig:Fits to Data} 
(\( (\alpha ,\beta ,\Lambda )=(1.76,0.386,0.789) \))
with \( \sigma _{N}=65.8 \) MeV. Clearly the curvature associated with the
chiral corrections at low quark mass is extremely important in the evaluation
of \( \sigma _{N} \).

In order to estimate the error in the extracted value of \( \sigma
_{N} \) we would need to have the full data set on a configuration by
configuration basis. As this is not available, 
the errors that we quote are naive estimates
only.  The extracted value of \( \sigma _{N} \) is very well
determined by the present data, the result being \( 54.6 \pm
2.0 \) MeV. Since the process of setting the physical mass scale via
the string tension is thought to have a systematic error of 10\%, one
might naively expect this to apply to \( \sigma _{N} \).  However,
{\em all} masses in the problem including the pion (or quark)
mass, as well as that of the nucleon, scale with the lattice parameter
\( a \).  It turns out that when one uses
Eq.~(\ref{eqn:FH-Commutator}) at the physical pion mass (which means a
slightly different value of \( \bar{m}a \) if \( a \) changes), the
value of \( \sigma _{N} \) is extremely stable. If, for example, one
raises the CP-PACS data by 15\% and the UKQCD data by 5\% (instead of
5\% and \( -5\% \), respectively) the value of \( \sigma _{N} \)
shifts from \( 54.6 \pm 2.0 \) to \( 55.2\pm 2.1 \) MeV. 
We present calculations in
Table \ref{table:Sigma Commutator Values} that show, for a variety of
scalings of the lattice data, how stable our results are.
\begin{table}
{\centering \begin{tabular}{|c|c||c|c|c|}
\hline 
\multicolumn{2}{|c||}{Scaling}&
\multicolumn{3}{|c|}{\( \sigma _{N} \) }\\
\hline 
{\tiny CP-PACS}&
{\tiny UKQCD}&
Dipole&
Sharp&
Cubic\\
\hline 
\hline 
5&
-5&
47.2\( \,\pm\, 1.8 \)&
54.6\( \,\pm\, 2.0 \)&
29.7\\
\hline 
10&
0&
48.1\( \,\pm\, 1.9 \)&
54.9\( \,\pm\, 2.0 \)&
28.6\\
\hline 
0&
-10&
45.4\( \,\pm\, 1.9 \)&
54.3\( \,\pm\, 1.9 \)&
31.0\\
\hline 
\end{tabular}\par}
\caption{Sigma Commutator Values. 
The \textbf{Dipole} and \textbf{Sharp} results were
calculated with our preferred form of \protect\( \alpha +\beta m_{\pi }^{2}+
\sigma _{NN}(\Lambda ,m_{\pi })+\sigma _{N\Delta }(\Lambda ,m_{\pi })\protect \)
with either a dipole form factor for the \protect\( N\pi \protect \)
vertex or a \protect\( \theta \protect \)-function.  The values of 
dipole parameter ($\Lambda_{D}$) were (1.225, 1.250, 1.175) GeV. The  
\textbf{Cubic} results are for the \protect\( \alpha +\beta m_{\pi
}^{2}+\gamma m_{\pi }^{3}\protect \) extrapolation
function, with \protect\( \gamma \) {\em unconstrained} by chiral
symmetry -- as explained in the text this produces an unreliable value
for \protect\( \sigma_{N} \).
\label{table:Sigma Commutator Values}}
\end{table}

The remaining issue, for the present data, is the model dependence associated
with the choice of a sharp cut-off in the pionic 
self-energies. Our investigations
in Ref.~\cite{Leinweber:1999ig} showed that Eq.~(\ref{eqn:Full-Eqn}) could
reproduce the dependence of \( M_{N} \) on \( m_{\pi }^{2} \) within the cloudy
bag model, and that it could also describe the dependence of pion self-energy
terms calculated with dipole form factors. 
Thus we believe that any model satisfying
the essential chiral constraints and fitting the lattice 
data should give essentially
the same answer. We checked this by numerically fitting the lattice data (solid
curve) with the form of Eq.~(\ref{eqn:Full-Eqn}) but with \( \sigma _{NN} \)
and \( \sigma _{N\Delta } \) calculated 
with dipole form factors of mass \( \Lambda _{D} \)
at all pion-baryon vertices.
Since the preferred phenomenological form of the $N\pi$ form factor
is a dipole, we regard the dipole result shown in the first line of
Table~\ref{table:Sigma Commutator Values} as our best estimate, namely
$\sigma_{N} = 47.2 \pm 1.8$ MeV with fit parameters \( (\alpha ,\beta
,\Lambda _{D})=(2.02,0.398,1.225) \). A remaining source of error
is that, although the lattice results were
calculated with an improved action, there still is an error associated
with the extrapolation to the infinite volume, continuum limit.
The importance of the inclusion
of the correct chiral behaviour is clearly seen by the fact that it increases
the value of the sigma commutator from the 30 MeV of the unconstrained
cubic fit to around 50 MeV.  

Clearly an enormous amount of work remains to be 
done before we will fully understand
the structure of the nucleon within QCD. It is vital that the rapid progress
on improved actions and faster computers continue and that 
we have three flavour
calculations within full QCD at masses as close as possible to the physical
quark masses.
Nevertheless, it is a remarkable result that the present lattice data for 
dynamical-fermion, 
two-flavour QCD, yields such a stable and 
accurate answer for the sigma commutator, an answer which is already within
the range of the experimental values. The implications of this result for models
of hadron structure need to be explored urgently.

One of us (SVW) would like to acknowledge helpful discussions with Tom Cohen
at an early stage of this work. We would also like acknowledge helpful
comments from Chris Allton, Craig Roberts and Robert Perry.
This work was supported in part by the Australian Research Council.

\bibliographystyle{utcaps}
\providecommand{\href}[2]{#2}\begingroup\raggedright\endgroup
\end{document}